\documentclass[pdflatex,sn-mathphys-num]{sn-jnl}

\usepackage{graphicx}%
\usepackage{multirow}%
\usepackage{amsmath,amssymb,amsfonts}%
\usepackage{amsthm}%
\usepackage{mathrsfs}%
\usepackage[title]{appendix}%
\usepackage{xcolor}%
\usepackage{textcomp}%
\usepackage{manyfoot}%
\usepackage{booktabs}%
\usepackage{algorithm}%
\usepackage{algorithmicx}%
\usepackage{algpseudocode}%
\usepackage{listings}%
\usepackage{xspace}%
\usepackage{tikz}%
\usetikzlibrary{calc,positioning}%
\usepackage{pdflscape}%

\theoremstyle{thmstyleone}%
\theoremstyle{thmstyletwo}%

\theoremstyle{thmstylethree}%

\def\pyt{\textsc{Pythia}\xspace}
\def\pyts{\textsc{Pythia}'s\xspace}
\def\jet{\textsc{Jetset}\xspace}
\def\ie{\textit{i.e.}\xspace}
\def\eg{\textit{e.g.}\xspace}
\def\qbar{\overline{q}}
\def\pbar{\overline{p}}

\begin{document}

\title[\pyt]{The \pyt Facility}


\author[1]{\fnm{Christian} \sur{Bierlich}}\email{christian.bierlich@fysik.lu.se}

\author*[1]{\fnm{Leif} \sur{Lönnblad}}\email{leif.lonnblad@fysik.lu.se}

\author[1]{\fnm{Torbjörn} \sur{Sjöstrand}}\email{torbjorn.sjostrand@fysik.lu.se}

\equalcont{These authors contributed equally to this work.}

\affil*[1]{\orgdiv{Department of Physics}, \orgname{Lund University}, \orgaddress{\street{Professorsgatan 1B}, \city{BOX~118}, \postcode{221 00}, \state{Lund}, \country{Sweden}}}




\abstract{The development and operation of large-scale particle physics facilities rely not only on accelerators and detectors, but also on sustained, high-precision simulation infrastructure. Originating in Lund in the late 1970s and continuously developed in Sweden for nearly five decades, \pyt has evolved into one of the most widely used Monte Carlo event generators in high-energy physics. Today it functions as a facility-scale software infrastructure underpinning the physics programmes of major international experiments, including those at the Large Hadron Collider, and plays a central role in validation, tuning, and uncertainty evaluation.

In this article, we present \pyt as a Swedish contribution to big science facilities. We outline its historical development, analyze its contemporary user base through citation and text-based studies, and map its integration across experimental frameworks, generator ecosystems, validation infrastructures, and emerging machine-learning workflows. These analyses show that \pyt is deeply embedded in global production chains and phenomenological studies, serving tens of thousands of researchers across multiple subfields.

We discuss the operational model and sustainability challenges associated with maintaining long-lived research software at facility scale. As particle physics moves toward the High-Luminosity LHC era and future facilities such as the EIC and FCC, continued investment in robust, interoperable simulation infrastructure remains essential.}

\keywords{Particle Physics, Software infrastructure, Monte Carlo simulations, Quantum Chromodynamics}



\maketitle

\tableofcontents
\newpage

\section{Introduction}\label{sec:intro}

Large-scale particle physics facilities are typically identified with
their physical infrastructure: accelerators, detectors, computing
centres, and the international collaborations that operate them.
However, the scientific output of such facilities also depends on
a less visible but equally critical layer of infrastructure:
theoretical modelling, simulation, validation, and analysis
frameworks that allow experimental data to be interpreted in terms of
fundamental physics. In this broader sense, modern particle physics
relies on a distributed ecosystem of software facilities that are not
bound to a geographical site.

\pyt \cite{Bierlich:2022pfr} is one such facility. It is not a
physical installation, but a software package providing a kind of
virtual-reality environment that scientists can visit to explore particle
physics scenarios and hopefully return wiser. With roots almost half a
century ago in Lund, Sweden, and with sustained leadership and
development anchored there ever since, \pyt for about 40 years has
been the most widely used program for particle physics modelling in
the world.

In this article, we present \pyt as a Swedish contribution to
big science facilities in the global sense. While not tied to a
single experimental site, \pyt has been, and continues to be,
integrated deeply into the physics programmes of major international
experiments. Its inception and development in Sweden has provided a stable and
long-term foundation for modelling high-energy collisions at
facilities such as PETRA, LEP, HERA, the Tevatron, the LHC, and upcoming
projects including the EIC and FCC studies. This Swedish contribution is not in the form of hardware components,
but rather of a sustained virtual infrastructure
that enables the extraction of physics from experimental data.

\begin{figure}
    \centering
    \includegraphics[width=0.70\linewidth]{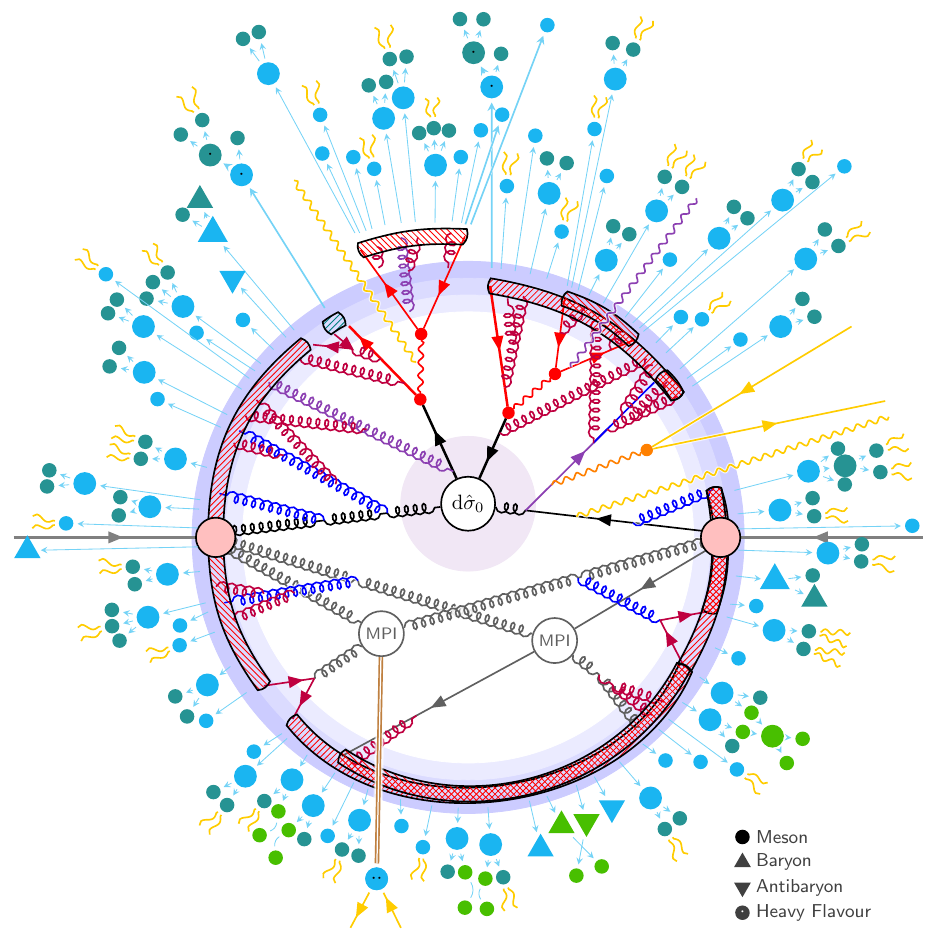}%
\includegraphics[width=0.30\linewidth]{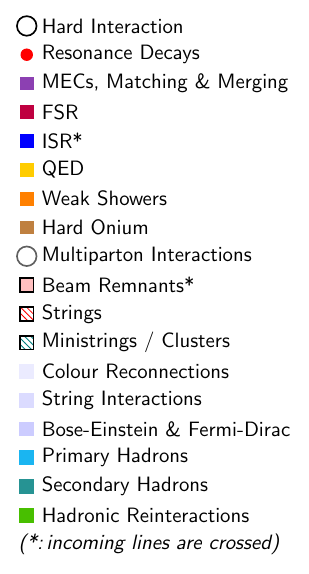}
    \caption{A schematic illustration of a particle collision,
    showing how different physics mechanisms contribute to the
    overall structure of the final event. Time is running outwards
    from the center of the figure.}
    \label{fig:event}
\end{figure}

More specifically, \pyt is a Monte Carlo event generator of
high-energy particle collisions. That is, it describes how the
collision between two incoming high-energy particles, say protons
like at the LHC at CERN, involves various processes that ultimately
can lead to hundreds of new particles being produced. In Fig.~\ref{fig:event} the structure of a simulated proton--proton collision is sketched, with all processes highlighted. All of these
processes are of a quantum mechanical nature, and random numbers
are used to pick between possible outcomes at every step on the way.
The rules of this game should ideally come from the Standard Model
(SM) of particle physics, which is consistent with all observations
to date, but its mathematical complexity defies complete,
exact solutions in most cases relevant to experimental studies. In some cases perturbation theory can be used to provide increasingly accurate approximations, in others not. The latter especially
applies to the QCD part of the SM, \ie the theory of strong interactions,
where QCD-inspired modelling is key to making any progress.

Thus, writing an event generator like \pyt is not only a matter of
programming skills, but even more one of having good physics ideas
on all of the processes involved in a particle collision. We typically
speak of a dozen different physics mechanisms that need to be modelled,
see Fig.~\ref{fig:event}, but each of these may be of considerable
complexity and need to be subdivided further into a multitude of
possible elements. In order to test experimentally what is going on,
it is also often useful to set up straw-man alternative models, or at
least provide free parameters within a model.

\begin{figure}
\centering
\begin{tikzpicture}[font=\small]

\coordinate (A) at (90:3.2);    
\coordinate (B) at (210:3.2);   
\coordinate (C) at (330:3.2);   
\coordinate (O) at (0,0);       

\draw[line width=0.8pt] (A) -- (B) -- (C) -- cycle;

\node[align=center, text width=3.6cm] at ($(A)+(0,0.55)$)
{\textbf{Theory}\\[-0.2em]
Phenomenology and\\ model development};

\node[align=center, text width=3.6cm] at ($(B)+(-0.85,-0.55)$)
{\textbf{Software}\\[-0.2em]
Engineering, interfaces\\ and performance};

\node[align=center, text width=3.6cm] at ($(C)+(0.85,-0.55)$)
{\textbf{Experiment}\\[-0.2em]
Users, training, support\\ and feedback};

\node[align=center, rotate=60] 
at ($(A)!0.5!(B)+(0.0,0.35)$)
{\small \textit{constraints}};

\node[align=center, rotate=0] 
at ($(B)!0.5!(C)+(0,-0.25)$)
{\small \textit{reproducibility}};

\node[align=center, rotate=-60] 
at ($(C)!0.5!(A)+(0.0,0.35)$)
{\small \textit{requirements}};

\node at (O) {\includegraphics[width=2.3cm]{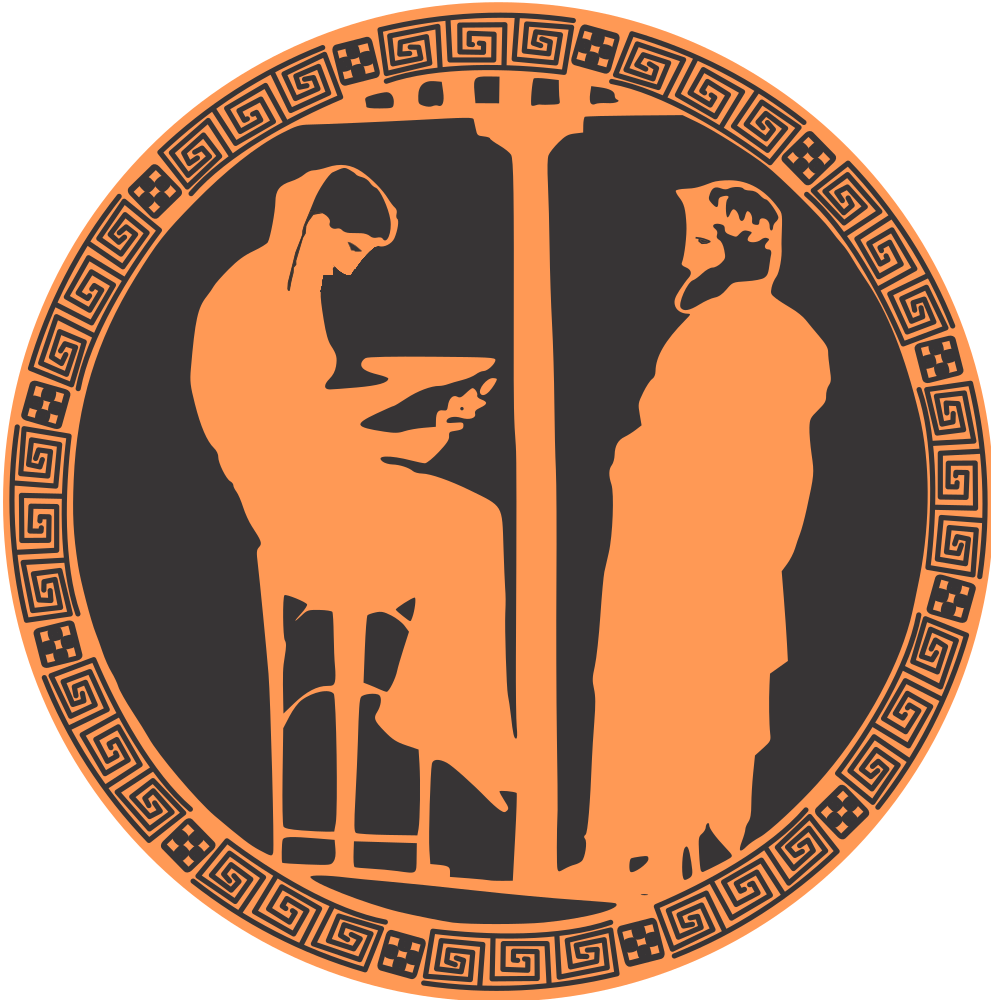}};

\end{tikzpicture}
\caption{The \pyt facility as the intersection of three coupled prongs:
theory-driven model development, software engineering, and continuous
interaction with the experimental user community.}
\label{fig:three-prong}
\end{figure}
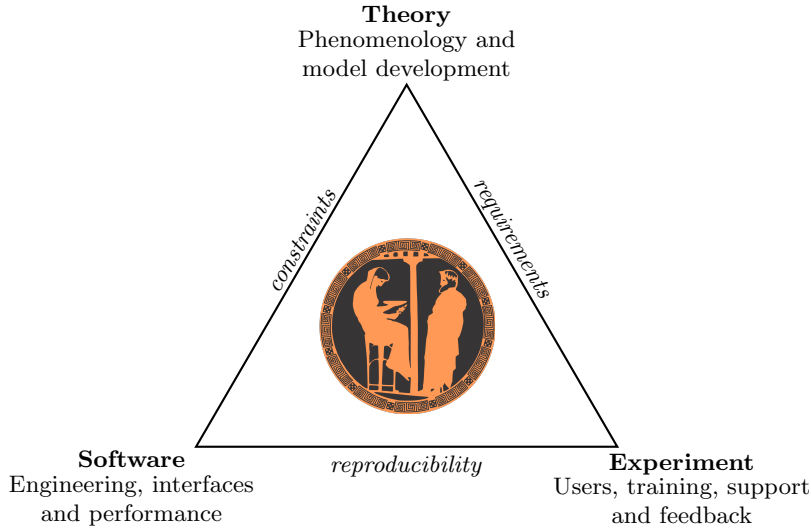

In practice, sustaining such modelling at the level required by modern collider facilities rests on three tightly coupled activities. The first is theory and phenomenology, where new physics models are formulated, extended, and confronted with data-driven tensions. The second is software engineering, where these models are turned into efficient, testable, and maintainable code, with stable interfaces to downstream tools and production environments. The third is the surrounding experimental community: continuous dialogue with users, training and support, and a feedback loop where analysis needs and emerging measurements shape development priorities. This three-prong nature is illustrated in Fig.~\ref{fig:three-prong}. The \pyt facility sits at the intersection of these three prongs, and its long-term impact comes precisely from maintaining their coupling rather than optimizing any one of them in isolation.

From the perspective of large experimental collaborations, event
generators such as \pyt function as integral components of the ecosystem. They are used in detector design studies, trigger optimization, background estimation, signal modelling,
systematic uncertainty evaluation, and long-term archival
reproducibility of analyses. They are embedded in production chains
that generate billions of simulated events and consume substantial
fractions of computing budgets. Their predictions must be continuously 
validated against data, and their evolution must be managed in a
controlled and documented manner. In this operational sense, an event
generator becomes a facility-like entity: it requires governance,
release management, validation infrastructure, user training, and
sustained scientific leadership.

The most vital part of the Swedish contribution to this ecosystem lies in this
long-term leadership. Since its inception in Lund in the late 1970s,
\pyt has been developed through successive generations of researchers,
with continuity of expertise and institutional memory that spans
multiple collider eras. This continuity has allowed the modelling of
QCD and related processes to evolve alongside experimental advances,
while preserving backward compatibility and reproducibility. It has
also positioned Sweden as a driving force in event-generator
development, influencing standards for event records, interfaces
between generators and experiments, and systematic tuning efforts
across the community.

In Section~\ref{sec:history} the historical origins and evolution
of \pyt are outlined. This involves hundreds of research articles,
and it would carry too far to cite them all. Instead we refer to
the comprehensive \pyt~6.4 \cite{Sjostrand:2006za} and
\pyt~8.3 \cite{Bierlich:2022pfr} overviews for details and references,
and to ref.~\cite{Sjostrand:2019zhc} for further history aspects.
Section~\ref{sec:today} gives an overview of the role \pyt currently
plays in particle physics (and beyond), and outlines some activities
too recent to have been included in ref.~\cite{Bierlich:2022pfr}.
Section~\ref{sec:future} offers a glimpse of future challenges.
Finally, some conclusions are drawn in Section~\ref{sec:conclusions}.

\section{History}\label{sec:history}

QCD, the theory of strong interactions, was presented in 1973. In it,
the interactions between quarks ($q$), antiquarks ($\qbar$) and gluons 
($g$) are uniquely defined, however in terms of so complicated
equations that no exact solution is known. Perturbation theory can 
be used to obtain increasingly precise answers in "hard" processes 
involving large momentum transfers, and lattice QCD correspondingly 
for static particle properties, but there is no obvious QCD formalism 
to address "soft" low-momentum-transfer processes. And yet it is the 
latter that are driving the multiparticle production at colliders. 

In the absence of even approximate QCD-derived approaches to soft
physics, QCD-inspired ones were developed. One of these was the
Lund string model, in which an assumed linear potential between
a $q_0$ and a $\qbar_0$ --- as favoured by lattice QCD ---
can be expressed in terms of a straight string stretched between 
the two. As $q_0$ and $\qbar_0$  move apart from a common 
creation vertex, energy flows into the string between the two,
and eventually the string can fragment by the production of new
$q_i\qbar_i$ pairs. The $q_i$ from one break can combine with 
the $\qbar_j$ from a neighbour break to form a $q_i\qbar_j$ meson. 
In this way the original pair fragments into a sequence of mesons.  

In 1978 a computer simulation of another model spurred the 
Lund seniors to assign two PhD students to code up this string
model. At their (partial) disposal was the main computer of
Lund University, one core, one MHz clock speed, and one Mbyte 
memory. Input by punched cards, output by line printer. In spite 
of all these limitations, it was still possible to implement model 
improvements and carry out studies not possible by paper and pen.  

The big breakthrough came when the model was extended to 
$q\qbar g$ systems, as created in the then new PETRA $e^+e^-$ 
collider at DESY in Hamburg. The Lund prediction here was that 
a string would be stretched from $q$ to $g$ and then on to
$\qbar$. Since there would be no direct connection between 
$q$ and $\qbar$, that angular region would have a depleted 
particle production relative to other models on the market. 
When this was confirmed by the JADE collaboration in 1980, it 
began a process where the Lund code came to dominate the studies 
for and at $e^+e^-$ colliders, including PEP and SLC at Stanford, 
TRISTAN in Japan and LEP at CERN, as the model evolved to meet
new demands.  

The code used for these studies, called \jet, consisted of two
components, one based on perturbation theory to set up the basic
collision process, $e^+e^- \to \gamma^*/Z^0 \to q\qbar /
q\qbar g / q\qbar g g / q\qbar q'\qbar'$, 
and another to carry out the subsequent string fragmentation, 
including decays of unstable particles. This modularity allowed 
new codes to implement other processes, but still make use of the 
string framework. This includes \textsc{Lepto} for $ep$ 
collisions, \pyt for $pp$ or $p\pbar$ ones, \textsc{Fritiof} 
for heavy-ion ones, \textsc{Ariadne} for dipole parton showers, 
and many more. Most of these have died out, for lack of support 
as students have moved on.

\begin{figure}
    \centering
    \includegraphics[width=0.95\linewidth]{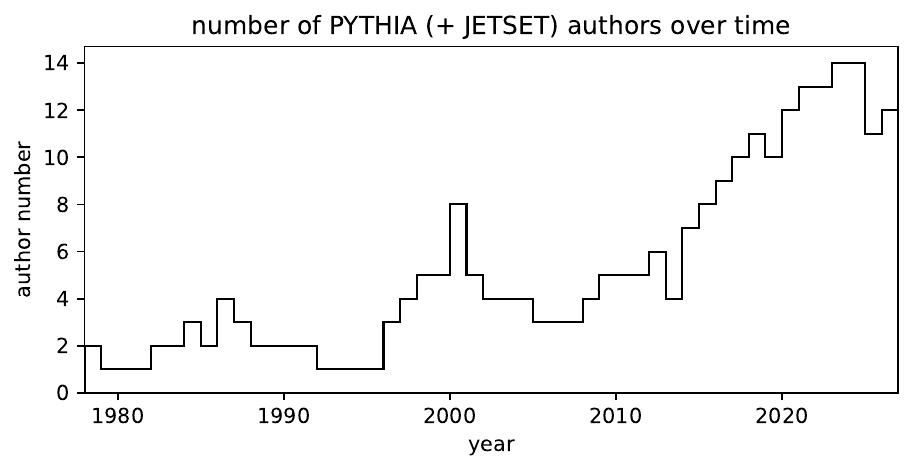}
\caption{The number of \pyt co-authors over time.}
\label{fig:authors}
\end{figure}

When the main \jet and \pyt authors both were postdocs in the US,
they became involved in the then burgeoning studies for the SSC
$pp$ collider. This led to an increased activity mainly related to 
the \pyt project, with the implementation of new processes,
of initial- and final-state parton showers, of multiparton
interactions, of colour reconnection, and some more. SSC was 
eventually cancelled, but the experience could be transferred 
directly to the LHC project at CERN. Once detector design 
studies began in earnest in 1990, \pyt quickly became the
main event generator, a role that it has maintained since. Notably,
the early designs of the ATLAS and CMS detectors were carefully
cross-checked against \pyt scenarios for Higgs production and decay
over a wide mass range. Not only for the Standard Model Higgs that
we now know agrees well with data, but also with a set of alternative
scenarios that were proposed at the time, and for many other processes
within and beyond the Standard Model.

In the early nineties only one author remained, so the \jet and \pyt
codes became increasingly coordinated, and were eventually joined
under the \pyt~6 label. But by then the group had started to increase
again. Over the subsequent years expansion of the code continued in 
various directions, with further processes, with improved parton showers, 
colour reconnection and other physics modelling, and with the addition 
of other beam types, notably photons. 

The CERN decision to move all computing from Fortran to C++ for the
LHC era also meant that \pyt had to be rewritten. The bulk of this job
was done in 2004--2007. At the same time the old Fortran version was 
maintained, and even continued to expand, and the last \pyt~6 update was 
released as late as 2013. To this day the Fortran code is still in use,
in communities where that language is the norm, but it is no longer 
supported.

\begin{figure}
    \centering
    \includegraphics[width=0.95\linewidth]{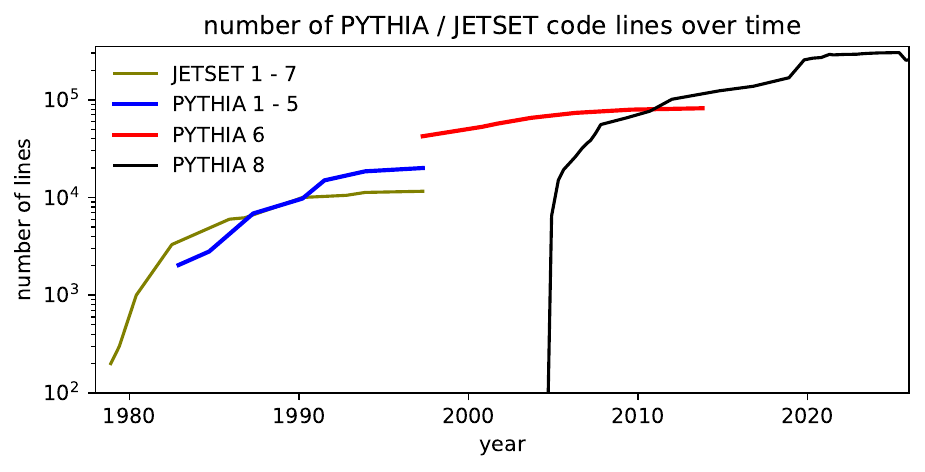}
    \caption{The size of the \pyt~code over time. The number of lines includes comments and blanks. Headers, source code and the examples main programs are included in the \pyt~8 count, but not data or documentation files.}
    \label{fig:programsize}
\end{figure}

The number of active code developers has increased in the years since 
the 2007 \pyt~8 release, see Fig.~\ref{fig:authors} for an overview. At the time of writing, there are 12 authors registered. The amount of code has also increased, see Fig.~\ref{fig:programsize}, where the transition between program versions are also represented. Since the first release in the \pyt~8 series, the code size has increased by a factor of five. Some of the later additions were less urgent parts of the Fortran code, that had to be left out in the first round for lack of time, but
most of the additions introduce new physics features.  Below we give a 
brief and non-exhaustive survey of such developments.   

The chance to start over from scratch allowed a clean-up of the code,
\eg with a new organizational structure. It notably included several 
interfaces to other codes in particle physics, extending and 
systematizing on what had been shoehorned into \pyt~6. One key example 
is that perturbative cross section expressions used to be typed by hand 
into \pyt, but by the turn of the millennium the automation of calculations 
could generate extremely long expressions for higher-order processes. 
It then made sense to let such codes, designed  to generate a handful 
of particles, provide the input to \pyt for the subsequent handling, 
on par with how internally implemented processes were generated.

Thus rather few matrix elements are hand-coded specifically for \pyt 
any longer, but instead imported from external generators when needed. 
But there are some exceptions, like charmonium and bottomonium 
production, notably colour octet such. Also, some BSM particles and 
processes have been added, \eg in the supersymmetry setup (including 
$R$-hadrons), and for contact interactions and Large Extra Dimensions.

Parton showers continue to play an important role. Besides the 
default shower methods, two other shower codes have been distributed
as part of \pyt, \textsc{Vincia} and \textsc{Dire}. The latter has
been orphaned and has been removed, but the former is a testbed 
for many new shower ideas towards improved precision, not only by
next-to-leading-log accuracy but also by a better and more predictable  
coverage of phase space. It also includes a full electroweak shower 
handling. The default shower has also seen some additions, such as
a dipole option for initial-state radiation, and the production of 
charmonium and bottomonium as part of the shower evolution. It has
also been extended to handle Beyond-the-Standard-Model (BSM) Hidden 
Valley scenarios, where the hidden quarks can both radiate and 
hadronize, with some particles potentially decaying back into our 
sector.

While parton showers can accurately describe parton emissions in the
soft and collinear limit, hard and wide angle emissions are generally
better handled by multi-leg matrix elements. Care must, however, be
taken when combining such matrix elements with parton showers, so that
one avoids double counting (or under counting) of emissions. \pyt was
the first generator that implemented a \textit{matching} of the
splitting probability in the shower to three-parton matrix elements to
get a smooth transition between them. Since
the beginning of the millennium, there has been several new techniques
developed to combine parton showers with higher order matrix elements
with more legs and loops, and \pyt today includes a wide
range of such techniques.

The string fragmentation framework remains at the center of \pyt,
essentially unchanged. Nevertheless, a number of variations have 
been introduced, in part in response to LHC data, and some may become 
new defaults in the future. Ropes, where several nearby strings 
combine into higher colour representations, offer an explanation 
for the enhanced strangeness production in high-multiplicity $pp$ 
collisions. An alternative is closepacking, where strings remain 
separate but obtain an increased string tension. Closepacking could 
also give rise to shoving, where strings receive a transverse 
motion that results in collective flow. Another potential flow 
mechanism is hadronic rescattering, where the space--time motion 
of the outgoing hadrons leads to collisions, usually at low energy.
In thermodynamical string fragmentation the particle composition and 
transverse momentum is driven by exponential suppression rather than 
Gaussian one. 

Colour reconnection of strings remains central. One important
development is the introduction of junction--antijunction formation,
where a junction corresponds to a Y-shaped combination of three 
colour strings. This gives a new mechanism for baryon production,
and explains the observed enhancement of $\Lambda_c$ at the LHC. 
New colour reconnection variations give an important source of top 
mass uncertainty, and are relevant \eg for $W^+W^-$ event shapes at 
$e^+e^-$ colliders.

In the handling of particle decays, the largest improvement is that
$\tau$ lepton production and decay is performed with full spin 
density matrices. 

Several different models for $pp$ total, elastic and differential 
cross sections have been added as options. Multiparton interactions 
have been implemented in diffractive systems, and hard diffraction
is handled with a dynamical rapidity gap survival from such 
interactions. 

\pyt/\jet was early on used also for collisions between heavy ions. In
particular, the Fritiof program developed in Lund, used the \jet
string fragmentation together with the so-called wounded nucleon model 
to simulate heavy ion collisions giving a very good description of the
experimental data available at the time. Since then, \pyt has included
its own heavy ion capability based on the \textsc{Angantyr} model. 
This model takes Fritiof as inspiration, but
rather than just having a single string per wounded nucleon, it
includes the full \pyt multiparton interaction machinery and parton
showers to model individual nucleon--nucleon subcollisions, making it
much better suited to model the much higher collisions energies
attainable at the Relativistic Heavy Ion Coller in Brookhaven and at
LHC.

Photoproduction, $\gamma\gamma$ and DIS (Deeply Inelastic Scattering) 
has also been implemented. The former can be applied \eg for UPC 
(Ultra-Peripheral Collisions) in heavy-ion collisions. But the future 
target is to build up this machinery for studies at the EIC 
(Electron--Ion Collider) planned in the US, where nuclear effects 
need to be included.

The evolution of cascades in matter is driven by the particles with 
largest momentum after each layer of interactions, which means that
the modelling of beam remnants is important, and needs to be
improved. This is also relevant for forward physics at the LHC, 
like future studies at the Forward Physics Facility.

\section{Today}\label{sec:today}

\pyt today is much more than just a program for simulating particle
collisions. It has gradually developed into an important software
infrastructure for high energy physics and beyond, and is a central
and indispensable component of the software ecosystem surrounding
modern collision experiments. It has a modular structure with well
defined interfaces for input, steering, and output. And while it is
still self-contained and can be used as a \textit{black box} producing
more or less realistic collision events, it is easy to break out parts of
it to be used within other applications, as well as using the full
infrastructure as a test bed for trying out new theoretical ideas for
parts of the full event simulation.

In this section we will try in some depth to analyse who is using the
\pyt facility and how. We will also briefly describe how the \pyt
infrastructure is maintained and by whom.

\subsection{User base}
\label{sec:user-base}
Since \pyt is released as Open Source software, downloadable without any registration, we do not keep track of its users. The most robust, systematic, public handle on its usage is the citations to the \pyt manual. Since its inception, \pyt manuals have accumulated more than 40,000 citations, from a large and diverse user base. For this contribution, we have analysed works citing the \pyt~8 manuals since 2018 \cite{Bierlich:2022pfr,Sjostrand:2014zea,Sjostrand:2007gs}, to give a current picture of the user base.

\begin{figure}[!htb]
    \centering
\includegraphics[width=0.75\linewidth]{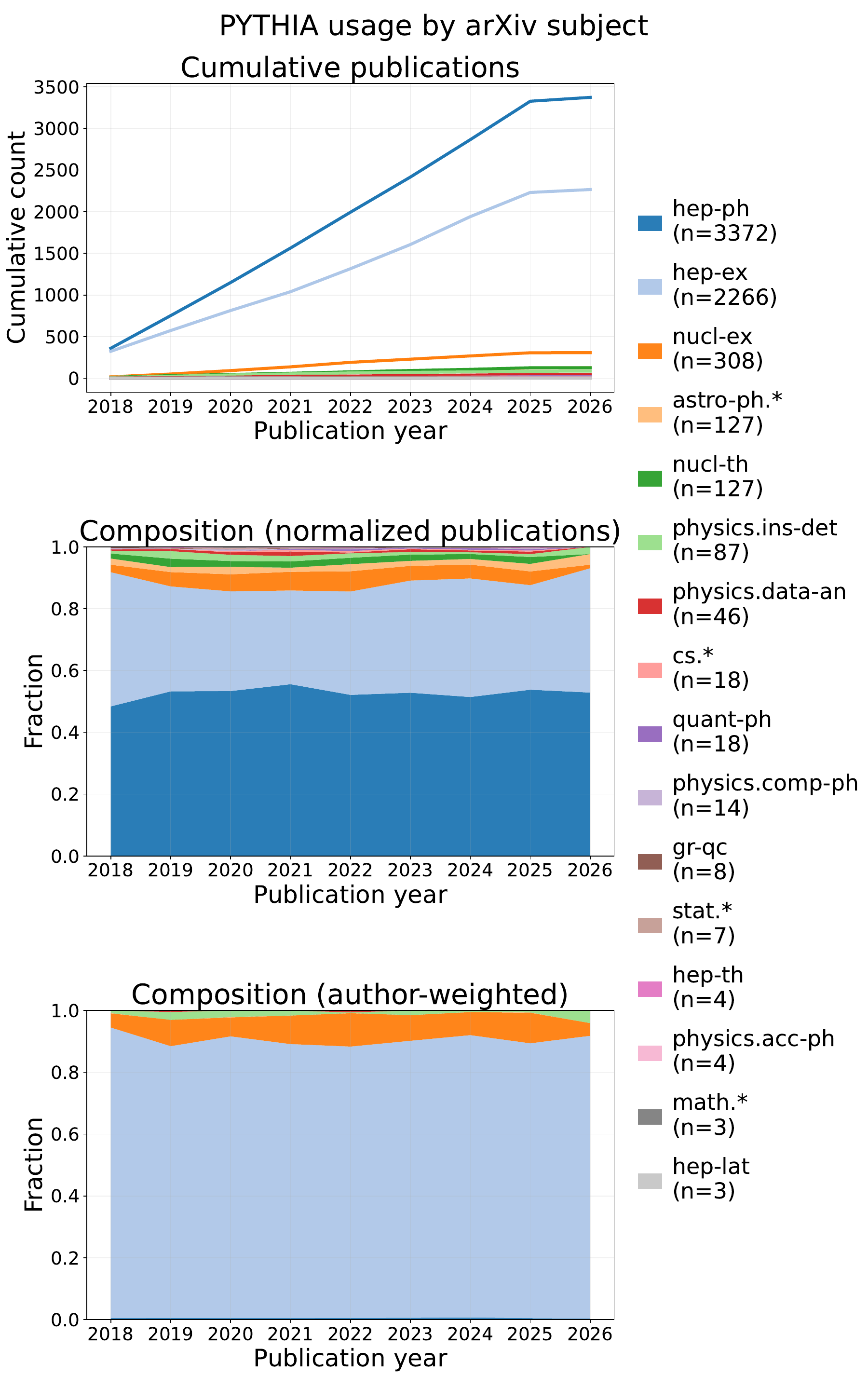}
    \caption{Overview of the \pyt user base based on works citing the \pyt~8 manuals since 2018, uploaded to arXiv. 
Left: cumulative citations by arXiv category. Center: normalized annual composition of publications. Right: normalized annual composition weighted by unique authors.}
    \label{fig:threepanel_arxiv}
\end{figure}

The three \pyt manuals in question have in total been cited approximately 10,000 times since 2018. Most citing works (6412 works out of 9641 works analysed) have been uploaded to the arXiv\footnote{The webpage \texttt{https://www.arxiv.org} is the main pre-print archive for high-energy physics.}, and are thus categorized by subject by its authors. This classification is, in Fig.~\ref{fig:threepanel_arxiv}, shown as function of time as (left) the cumulative citation count by subject, (center) the normalized composition of publications per year, and (right) the corresponding normalized distribution weighted by unique authors. The corpus had a total of 47,295 unique authors. This is taken as a proxy of the publishing user base, \ie the number of scientists publishing papers depending on \pyt. While some papers in the corpus are citing \pyt without actually using it, there are, on the other hand,  scientific papers using \pyt indirectly linked through other software, without citing \pyt.

When normalized by unique authors, it is clear that while the majority of papers citing \pyt are phenomenological studies classified \texttt{hep-ph}, the nature of high-energy physics, with very large experimental collaborations, make the experiments dominate in terms of individual users. Another interesting point to take away from the classification analysis is that \pyt experiences very diverse use, with works classified broadly across several physics domains, as well as computer science and statistics.

Within the sub-fields of high-energy physics, however, arXiv classifications or self-assigned keywords are too broad to allow a meaningful deeper characterization of the user communities. We therefore performed a dedicated text-based analysis of the full corpus of citing works.

For each paper, the full available text was processed, including title, abstract, and main body where accessible. From this material we constructed a numerical representation of each document. The analysis proceeded in two complementary steps.

First, we defined a set of domains corresponding to established research directions within high-energy physics. For each domain, a set of characteristic keywords was specified by us. Papers were scored according to the presence of these keywords in the full text, providing an initial classification.

Second, each document was converted into a numerical vector representation (an embedding) that captures its overall scientific content. This was done by counting how frequently different terms appear in the text. Each word contributes to one component of a high-dimensional vector according to a fixed, deterministic hashing scheme. Words that appear many times in a paper contribute more strongly than rare words, but the contribution grows only logarithmically with frequency so that very common words do not dominate. The resulting vector is normalized to unit length.

In this representation, two papers that use similar terminology and discuss similar physics mechanisms will produce vectors that point in similar directions. Their similarity can therefore be quantified by the cosine of the angle between the two vectors (cosine similarity). This allows us to
group papers according to overall thematic similarity, even if they do not share exactly the same predefined keywords.

\begin{figure}[!htb]
    \centering
\includegraphics[width=0.75\linewidth]{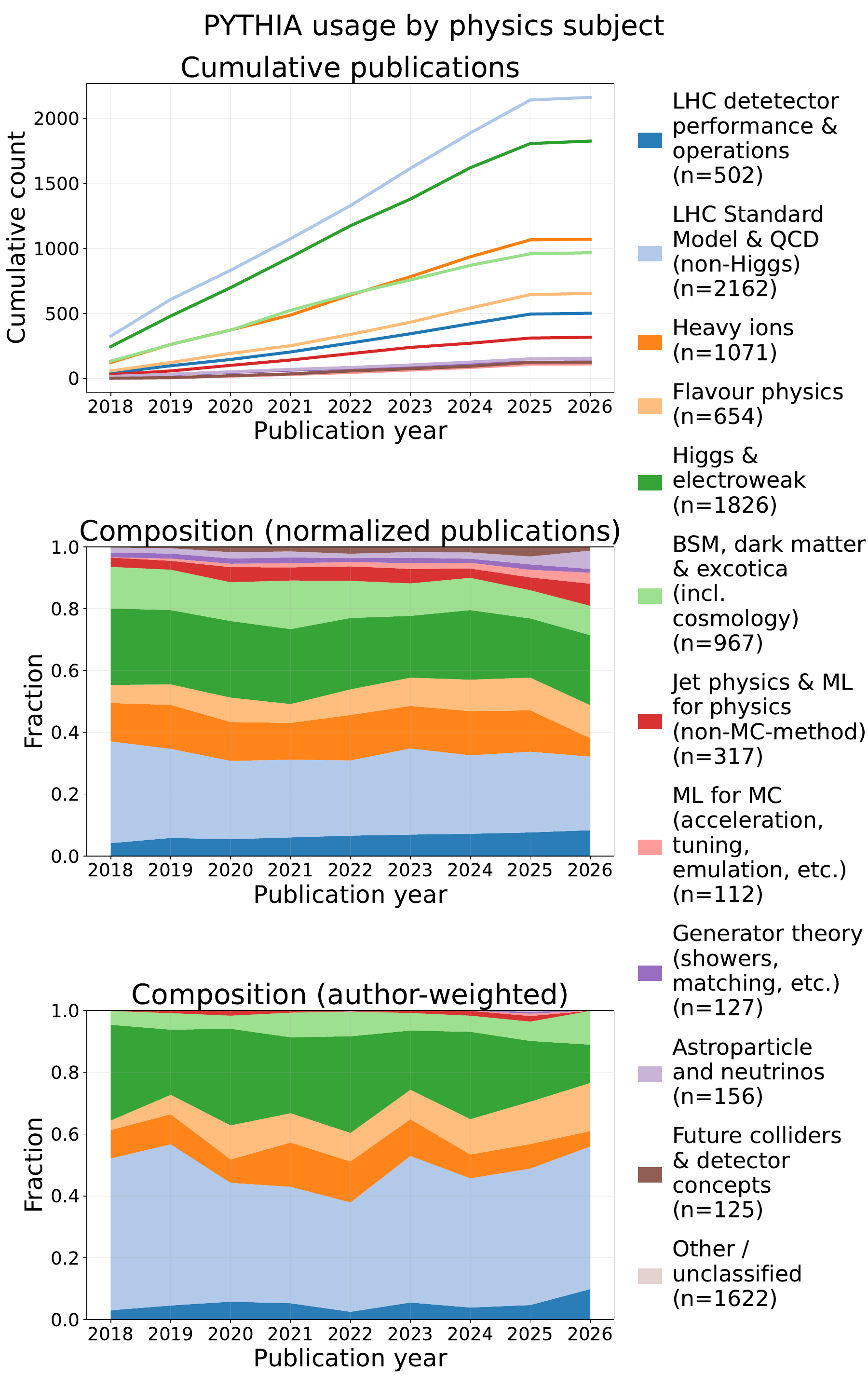}
    \caption{Overview of the \pyt user base based on works citing the \pyt~8 manuals since 2018. 
Left: cumulative citations by research domain, illustrating the dominant role of the LHC experimental program. Center: normalized annual composition of publications. Right: normalized annual composition of unique authors.}
    \label{fig:threepanel}
\end{figure}

Papers that could not be unambiguously assigned based on keywords alone were compared to the average vector (centroid) of already well-identified papers in each domain. If a document was sufficiently similar to one of these centroids, it was assigned to that domain. This refinement step was iterated until stable assignments were obtained. The final categorization was validated by manual inspection of representative papers in each group. The outcome of this analysis is summarized in Fig.~\ref{fig:threepanel}. 

This classification gives are more detailed overview of \pyt usage within high-energy physics. Usage is mainly distributed among the main research lines at LHC: Standard Model and QCD physics (light blue), heavy-ion physics (orange), flavour physics (light orange), Higgs and electroweak physics (dark green) and BSM (light green). Of these areas, Sweden has strong experimental and/or phenomenological involvement in four out of five through both experimental groups (\eg ATLAS, ALICE, LDMX) and theory groups at several universities. These groups are  using \pyt in a range of configurations: as a full event generator for baseline physics modelling; as part of analysis chains for acceptance and
systematics studies; and as a reference model to compare with data in measurements and searches. The heavy-ion and the flavour physics communities often use \pyt as part of a more specialized simulation chain, where it interfaces with other programs replacing parts of the model with other, context-dependent, tools.

Smaller and distinct user communities, such as \eg astroparticle physics and generator theory are also visible in this analysis, and as we will discuss later, distinct communities require different community efforts on our side. Finally, we note that \pyt is also well used in more future-pointing efforts, both future collider studies and the emerging field of machine learning for high-energy physics. 

\begin{figure}[!htb]
    \centering
\includegraphics[width=0.75\linewidth]{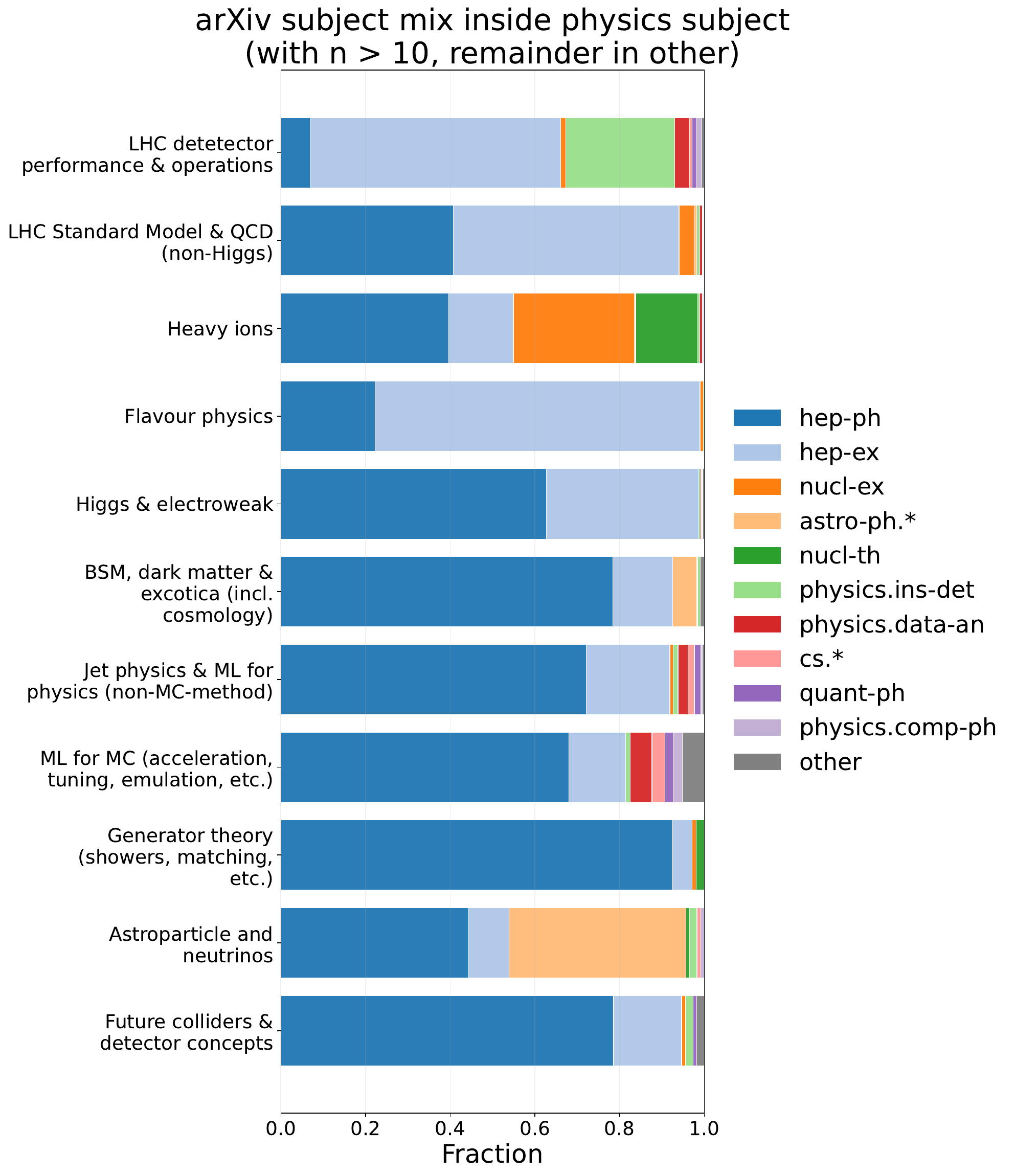}
    \caption{Physics subject breakdown in arXiv categories, showing how large a fraction of different physics subjects consist of different arXiv categories.}
    \label{fig:subject-breakdown}
\end{figure}

Finally, in Fig.~\ref{fig:subject-breakdown}, we cross-reference the two analyses above, to show how large a fraction of the physics subjects belongs to different author-assigned arXiv subjects.

Apart from the two obvious outliers in \texttt{exp/ph} split (LHC detector performance \& operations and Generator theory), all categories see a roughly even divide between experimental and phenomenological studies. This confirms that \pyt is being used in a broad sense among all categories, making the user groups even in each subject category quite diverse. 

This mixture implies that support cannot be optimized for a single ‘typical’ workflow. Stable production interfaces and rapid-prototyping modularity must coexist. The large LHC collaborations rely on long-term stability, careful validation, and predictable evolution of physics models and interfaces, across all physics subjects. They also require direct and sustained contact between \pyt authors and experiment software and analysis groups, to ensure that critical features and fixes are implemented in a way that is compatible with large-scale production workflows. At the same time, the decision and integration cycles in such environments can be slow, and support structures must be designed with that reality in mind.

Smaller collaborations and method-focused groups typically have a different mode of interaction. They often need more hands-on support, short feedback loops, and easy access to expert knowledge about both physics modelling and technical integration. For these users, a modular design that enables rapid turnaround on new ideas, together with clear paths for asking questions and receiving guidance from the authors, is particularly important.

In the following sections we discuss how the \pyt facility engages with these communities in practice, and how a more systematic support and training effort can help serve both the large, stability-driven user base and the smaller, fast-moving groups.

\subsection{Adoption rate by large experiments}

Since the large LHC collaborations dominate the author-weighted citation curves, it is useful to summarize the fraction of their published papers that explicitly cite \pyt. This is shown in Table~\ref{tab:citation_fraction}. The numbers reported there are the mean yearly explicit citation fractions over 2021--2026, where 2026 is included as a partial year.

\begin{table}[htbp]
\centering
\caption{Mean yearly explicit \pyt citation fraction from 2021 through June 9, 2026.}
\label{tab:citation_fraction}
\begin{tabular}{lc}
\hline
Experiment / selection & Mean citation fraction \\
\hline
ATLAS (\texttt{hep-ex}) & 0.913 \\
CMS (\texttt{hep-ex}) & 0.919 \\
ALICE (\texttt{hep-ex} + \texttt{nucl-ex}) & 0.565 \\
LHCb (\texttt{hep-ex}) & 0.900 \\
\hline
\end{tabular}
\end{table}

The citation-fraction analysis measures the fraction of papers published by the four experiments in 2021--2026 that cite one of the core \pyt~8 reference papers. We restrict the discussion to this narrower interval because some experiments were still transitioning from \pyt~6 to \pyt~8 before 2021. Overall, the explicit citation rate is high: three of the four experiments explicitly cite \pyt in $\geq 90\%$ of their publications. The clear outlier is ALICE.

At first sight this may appear unsurprising, since ALICE is focused on heavy-ion physics, which lies somewhat outside the traditional core use case of \pyt. However, as discussed later in Sec.~\ref{sec:ecosystem}, fields outside traditional high-energy collider phenomenology often use \pyt as part of a broader simulation chain, sometimes without a direct citation to the core \pyt references. To probe this possibility, we inspected five randomly selected ALICE publications \cite{ALICE:2018cqy,ALICE:2020wvi,ALICE:2022veq,ALICE:2024say,ALICE:2025ygv} from the subset that do not cite \pyt.

Four of the five papers are ion--ion measurements, while the fifth is a proton--proton study. The proton--proton paper explicitly mentions \pyt~8.2 but does not provide a citation to the core \pyt references. Three of the four heavy-ion papers cite \textsc{Hijing} \cite{Gyulassy:1994ew}, a generator framework with substantial  dependence on \pyt. The remaining paper mentions Monte Carlo background simulation without a clear citation trail.

This small spot check suggests that the ALICE citation fraction should be interpreted as a lower bound on actual \pyt usage, rather than as a direct measure of adoption. In other words, the explicit citation rate is lower in ALICE, but the true level of direct or indirect \pyt usage may be substantially higher than the citation numbers alone imply.

\subsection{Operations model}
\label{sec:operations-model}

Currently, there are 12 authors listed on the
\href{https://pythia.org}{pythia.org} website, a number that may seem small,
but it is not dissimilar to other similar sized research software. It
should be noted, however, that only three of the authors are tenured
senior researchers with \pyt as their main activity, whereof one is
closing in on retirement. The other authors typically have their main
activity in other collaborations, or are PhD students or Postdocs.

Just like the \textit{Lund string fragmentation} model is still the centre
piece of the \pyt program, Lund University is still the core and the
home of the \pyt collaboration. The collaboration is, however, truly
international and includes authors in other European universities, as
well as in institutes in the US, India and Australia.

The collaboration is led by board consisting of a Spokesperson, the
Technical Lead and the Release Manager, together with two more
authors, all of which are typically elected during the yearly
in-person \pyt Week. While \emph{authorship} is fairly well defined
\cite{agreement}, collaboration \emph{membership} is more informal and
decided by the board on a case-by-case basis.

The general aim is to have around three releases per year, but there
is no fixed release schedule. Instead, when a release should be made
and what goes into it, is discussed in the monthly (on-line)
collaboration meetings. These meetings also involve discussions about
issues that have been reported through the \texttt{gitlab} issue desk
\cite{gitlab} or to the contact persons assigned to the major
experiments, as well as more internal issues and progress reports on
physics and code.

While the Authors are responsible for the maintenance and development
of the code, input from the user community is welcome. The model
adapted by the collaboration for this is a bit more restrictive than
the completely open model where anyone can create a merge request on
\texttt{gitlab}. Instead the development branches of \pyt are private
to the collaboration members, and contributions from other users are
typically included by one of the authors. Contributions are properly
credited in the update notes and in the on-line manual.

\label{sec:licensing}
\pyt is licensed under the \textit{GNU GPL v2 or later} and the use of
the code within other programs is allowed and also encouraged, as will
be discussed below in section \ref{sec:ecosystem}. Modifications to
the code is also allowed, but is discouraged due to the obvious
maintenance issues. Instead, \pyt is highly modular, and allows users
to develop their own modules to change the behaviour according to
their needs. There is also a system of so-called \texttt{UserHooks}
where a user can go in at predetermined places and change the course
of the event being generated.

\subsection{Community engagement, summer schools, tutorials}
\label{sec:community}
Clearly, a significant fraction of users are members of large
experimental collaborations at the LHC and elsewhere. These
collaborations normally have their own software infrastructure for
interfacing to \pyt and other event generators (see e.g.\ table
\ref{table:embedding-experimental} below), and they typically then
also deal with user interactions themselves. In this way the generator
authors are somewhat insulated from common and trivially solvable
issues that a user may encounter. For this to function, however, there
need to be members of the collaborations that are experts. Making sure
that such expert users are available has since 20 years been one of
the cornerstones in the activities of the MCnet Collaboration
\cite{MCnet}.

MCnet was started by \pyt authors, together with authors of the other
two general purpose event generators, \textsc{Herwig} and
\textsc{Sherpa}, as a Marie S{k\l}odowska Curie doctoral network with
funding from the European Union. Besides training a new generation of
event generator authors, the goal of this network was also to train
the user community. This was done by arranging yearly summer schools
on the ``Physics and techniques of Event Generators'' with hands-on
tutorials, but also by funding so-called short term studentships,
where any PhD student could apply come to one of the network nodes for
3--6 months to work on a project together with event generator
authors. Over the years MCnet has organised more than 20 schools
(whereof two in Lund) and has hosted almost 80 short-term students
(whereof 19 in Lund). After three almost consecutive four-year grants
from the EU, MCnet is now not funded, but nevertheless continues its
annual summer schools and short-term studentships with support from
the LHC Physics Center at CERN.

Not all users are part of large experimental communities, and the
interaction with these are mainly maintained through our public
\texttt{gitlab} repository \cite{gitlab}, which includes an active
issue desk. There we also maintain a number of tutorials on general
and specialised usage of \pyt. These were originally designed for
summer schools, and we frequently use these in summer schools arranged
by MCnet and others. These tutorials are also the base for the use of
\pyt in undergraduate teaching, where they are used as an important
bridge between the often very abstract concepts taught in theoretical
courses and the very hands-on approach of experimental
courses. Finally the tutorials may also be used as self-study
material, which not least important for students in developing
countries where the expertise among teachers and the computer power
may be lacking. Even with a modest laptop with limited internet
connection, anyone can download the \pyt code and a tutorial and
produce their first events within an hour or two.

\subsection{Ecosystem: Usage by other codes} 
\label{sec:ecosystem}

Beyond its direct user base, \pyt plays a central role in the wider high-energy physics software ecosystem. In this context, \pyt is not used directly by an end user to generate and analyze events, but is instead invoked as a dependency within a larger software framework. By ``usage by other codes'' we thus mean independently developed software frameworks -- either publicly released or internally maintained within large collaborations -- that depend on \pyt as part of their computational workflow. This excludes ordinary analysis scripts or private studies, and focuses instead on structured software products with sustained development and distribution.

The scope of such dependencies is broad and spans multiple modes of integration. For clarity, we distinguish between (i) runtime physics integration (Sec.~\ref{sec:runtime-integration}), (ii) embedding in experimental analysis frameworks (Sec.~\ref{sec:analysis-frameworks}), (iii) validation and tuning infrastructures (Sec.~\ref{sec:validation-infra}), and (iv) machine-learning applications (Sec.~\ref{sec:ml-applications}). Educational activities and community training efforts are discussed separately in Sec.~\ref{sec:community}, while standardization and interface development are treated separately in Sec.~\ref{sec:standards}.

\subsubsection{Runtime physics integration}
\label{sec:runtime-integration}
A large number of simulation codes integrate \pyt at runtime to provide one or more components of the event-generation chain. The depth of this integration varies: in some cases \pyt is an optional backend; in others it serves as the default hard-process and shower generator within a modular environment; and in some cases it is structurally embedded and not easily replaceable. 

The open-source license of \pyt (see Sec.~\ref{sec:licensing}) permits modification, redistribution, and linking without formal coordination with the original authors. While many integrations are known to us through collaboration and community interaction, the full set of downstream uses is difficult to enumerate. In Table~\ref{table:runtime-integration} we summarize published simulation codes known by us, by physics topic. Sometimes \pyt is only integrated to do string fragmentation, sometimes to do everything but for one replaced step in the normal flow. To illustrate the diversity of uses inside this category, we mention three different examples from Table~\ref{table:runtime-integration}, each showcasing a different way of using \pyt.

\begin{table}
\renewcommand{\arraystretch}{1.15}
\begin{tabular}{p{0.33\textwidth} p{0.62\textwidth}}
\toprule
\textbf{Domain} & \textbf{Tool} \\
\midrule
General purpose generators & 
\textsc{Sherpa} \cite{Sherpa:2024mfk}, 
\textsc{Herwig} \cite{Bellm:2025pcw} \\
\addlinespace[0.4em]
Matrix elements: LHC & 
\textsc{MadGraph\_aMC@NLO} \cite{Alwall:2014hca},
\textsc{PowHeg} \cite{Frixione:2007vw}, 
\textsc{CalcHEP} \cite{Belyaev:2012qa}, 
\textsc{AlpGen} \cite{Mangano:2002ea}, 
\textsc{Helac-Onia} \cite{Shao:2015vga} \\
\addlinespace[0.4em]
Resummation calculations: LHC & 
\textsc{Geneva} \cite{Alioli:2012fc}, 
\textsc{MiNLO} \cite{Hamilton:2012np} (+ extensions), 
\textsc{HEJ} \cite{Andersen:2017sht} \\
\addlinespace[0.4em]
Matrix elements: FCC-$ee$, LINACS \&
muon colliders & 
\textsc{Whizard}  \cite{Kilian:2007gr}
(and \textsc{MadGraph} also here) \\
\addlinespace[0.4em]
Mini black holes &
\textsc{TrueNoir}  \cite{Landsberg:2003br},
\textsc{Charybdis} \cite{Harris:2003db},
\textsc{Catfish} \cite{Cavaglia:2006uk},
\textsc{BlackMax} \cite{Dai:2009by},
\textsc{QBH} \cite{Gingrich:2009da} \\
\addlinespace[0.4em]
Parton shower programs & 
\textsc{Ariadne} \cite{Lonnblad:1992tz},
\textsc{NLLjet} \cite{Kato:1990as}, 
\textsc{Deductor} \cite{Nagy:2017dxh},
\textsc{Cascade} \cite{CASCADE:2010clj}, 
\textsc{Panscales} \cite{vanBeekveld:2023ivn}, 
\textsc{Alaric} \cite{Herren:2022jej} \\
\addlinespace[0.4em]
Deeply Inelastic Scattering & 
\textsc{Lepto} \cite{Ingelman:1996mq},
\textsc{RapGap} \cite{Jung:1993gf},
\textsc{PolDis} \cite{Bravar:1997hs},
\textsc{Django} \cite{Charchula:1994kf} \\
\addlinespace[0.4em]
DIS on nuclear targets & 
\textsc{Beagle} \cite{Chang:2022hkt} \\
\addlinespace[0.4em]
Heavy Ion collisions & 
\textsc{Fritiof} \cite{Nilsson-Almqvist:1986ast},
\textsc{Hijing} \cite{Gyulassy:1994ew} 
(and \textsc{Hijing++} \cite{Biro:2019ijx}),
\textsc{AMPT} \cite{Lin:2004en}, 
\textsc{UrQMD} \cite{Petersen:2008kb}, 
\textsc{Jewel} \cite{Zapp:2013vla}, 
\textsc{Q-Pythia} \cite{Armesto:2009fj},
\textsc{Smash} \cite{Sciarra:2024gcz},
\textsc{Jetscape} \cite{Putschke:2019yrg},
\textsc{Paciae} \cite{Lei:2024kam},
\textsc{HydJet++} \cite{Lokhtin:2008xi}, 
\textsc{PyQuen} \cite{Lokhtin:2005px}, 
\textsc{Dipsy} \cite{Flensburg:2011kk}, 
\textsc{YaJEM} \cite{Renk:2009hv} \\
\addlinespace[0.4em]
Photon-initiated processes & 
\textsc{STARlight} \cite{Klein:2016yzr},
\textsc{SuperChic} \cite{Harland-Lang:2018iur},
\textsc{CepGen} \cite{Forthomme:2018ecc},
\textsc{UPCgen} \cite{Burmasov:2021phy},
\textsc{eHIJING} \cite{Ke:2023xeo} \\
\addlinespace[0.4em]
Cosmic rays & 
\textsc{DPMjet} \cite{Roesler:2000he}, 
\textsc{Sibyll} \cite{Riehn:2019jet}\\
\addlinespace[0.4em]
Astroparticle \& Dark Matter & 
\textsc{Gambit} \cite{Bloor:2021gtp},
\textsc{DarkCast} \cite{Ilten:2018crw},
\textsc{DarkSUSY} \cite{Bringmann:2018lay} \\
\addlinespace[0.4em]
Neutrino beams &
\textsc{Genie} \cite{Andreopoulos:2009rq},
\textsc{Neut} \cite{Hayato:2021heg}, 
\textsc{GiBUU} \cite{Buss:2011mx}, 
\textsc{NuWro} \cite{Golan:2012rfa}, 
\textsc{Negn} \cite{Autiero:2005ve}, 
\textsc{Nuance} \cite{Casper:2002sd} \\
\addlinespace[0.4em]
B-hadron decays &
\textsc{Evtgen} \cite{Lange:2001uf} \\
\addlinespace[0.4em]
Detector simulation &
\textsc{Geant4} \cite{GEANT4:2002zbu}, 
\textsc{Fluka} \cite{Ballarini:2024isa}.
\textsc{Delphes} \cite{deFavereau:2013fsa},
\textsc{LArSoft} \cite{Church:2013hea} \\
\bottomrule
\end{tabular}
\vspace{0.5em}
\caption{Representative published frameworks known to integrate \pyt. The depth and architectural nature of this integration vary considerably and are not exhaustively classified here. We have restricted to one key reference per program, but most have evolved over the years, as reflected in long chains of relevant articles.} 
\label{table:runtime-integration}
\end{table}

\begin{description}
	\item[\textsc{Sherpa}] is a general-purpose event generator for LHC physics. It can optionally interface to \pyt to use the Lund string hadronization model instead of its native cluster model. This provides a mechanism for systematic studies of hadronization uncertainties. The integration is modular, and \textsc{Sherpa} can operate independently of \pyt.

	\item[\textsc{Jewel}] is a specialized event generator for jet quenching in heavy-ion physics. It is constructed as a set of modifications to the \pyt~6 code base, primarily altering the parton shower to include medium-induced effects. As such, \textsc{Jewel} is tightly coupled to \pyt and cannot remove this dependency.

	\textsc{Jetscape} is a modular framework for heavy-ion event generation. Within this architecture, \pyt is used as the default hard-process generator and vacuum shower, and as an option for hadronization. Event generators constructed within the \textsc{Jetscape} framework therefore inherit this dependency, although they are not listed individually in Table~\ref{table:runtime-integration}. While alternative components could in principle be substituted, this would require significant modification of the framework structure.

\end{description}

These examples illustrate that \pyt frequently functions as a foundational physics backend within larger simulation architectures, with roles ranging from optional module to structurally embedded core.

\subsubsection{Embedding in experimental analysis frameworks}
\label{sec:analysis-frameworks}
In addition to dedicated generator frameworks, \pyt is embedded within large experimental software environments. These frameworks integrate \pyt into detector simulation, trigger studies, and large-scale production chains. The typical work-flow in a production chain is that a central group inside the experiment is responsible for generation of samples on a large scale, which is then distributed to members of the collaboration. This work-flow can also be found in parallel initiatives such as CERN Open Data \cite{CarreraJarrin:2021soc}, where simulated events by \pyt are published to the public, matched to public data sets.

While dependencies in experimental frameworks are often modular and configurable, in practice they represent long-term operational integration within major collaborations. These frameworks are primarily developed and maintained within the experimental collaborations, although many components are now publicly accessible. However, this use of \pyt is, as shown in the analysis in Sec.~\ref{sec:user-base}, the largest user base of \pyt by author.

In Table~\ref{table:embedding-experimental} we list the main long-term embedded frameworks, used by large collaborations, where integration is documented and a citable source is available. However, \pyt is embedded in several smaller or emerging frameworks besides those, for example used by COMPASS, IceCube and EIC collaborations. In such cases, a citing publication is not always available, for example in the case of the \texttt{eic-shell} tool \cite{eic-shell} by the \textsc{ePIC} collaboration, which nevertheless uses \pyt~8 heavily.

\begin{table}
\renewcommand{\arraystretch}{1.15}
\begin{tabular}{lcr}
\toprule
\textbf{Facility} & \textbf{Experiment} & \textbf{Framework} \\
\midrule
LHC & ATLAS & \textsc{Athena} \cite{Calafiura:2005zz}, \textsc{AthenaMT} \cite{ATLAS:2017vbf} \\
& CMS & \textsc{CMSSW} \cite{Jones:2015soc}\\
& ALICE & \textsc{AliRoot} \cite{Brun:2003vw}, \textsc{$O^2$} \cite{Buncic:2015ari}\\
& LHCb & \textsc{GAUSS} (\textsc{GAUDI}) \cite{Barrand:2001ny} \\
KEK & Belle & \textsc{BASF} \cite{Itoh:1997st}\\
& Belle II & \textsc{basf2} \cite{Gelb:2018agf} \\
Fermilab & DUNE & \textsc{LArSoft} \cite{Church:2013hea}\\
RHIC & sPHENIX & \textsc{Fun4All} \cite{Pinkenburg:2011zza}\\
& STAR & \textsc{BFC} \cite{Pruneau:2005zz}\\
FCC & & \textsc{Key4hep} \cite{Carceller:2025ydg} \\
Cosmic Rays & Auger (and others) & \textsc{Corsika} \cite{CORSIKA:2025ops} \\
\bottomrule
\end{tabular}
\vspace{0.5em}
\caption{Representative experimental frameworks or analysis frameworks known to integrate \pyt.}
\label{table:embedding-experimental}
\end{table}

In contrast to more specialized generators, \pyt is frequently employed both for signal modeling and for inclusive background simulation in large-scale production campaigns.

\subsubsection{Validation and tuning infrastructures}
\label{sec:validation-infra}
Beyond operational embedding in experimental analysis and production chains, \pyt is integrated into a broader validation and tuning infrastructure spanning experiments and theoretical developments. These frameworks provide standardized event representations, reproducible analysis environments, and systematic calibration of model parameters.

A first layer consists of common data formats and analysis backbones enabling interoperability between generators and experiments. The \textsc{HepMC} event record~\cite{Dobbs:2001ck} defines a generator-independent format for simulated events, allowing \pyt output to be processed within experimental and phenomenological workflows. The \textsc{ROOT} framework~\cite{Brun:1997pa} provides the dominant environment for histogramming, statistical analysis, and visualization, underpinning many validation and detector studies involving \pyt.

Generator validation against published measurements is implemented in the \textsc{Rivet} framework~\cite{Bierlich:2019rhm}, which encodes analysis routines in a generator-independent manner. \textsc{Rivet} enables systematic comparison of \pyt predictions to data across a wide range of observables while preserving analysis definitions independently of specific generator implementations.

Dedicated tuning frameworks build on these interoperability layers to calibrate model parameters to data. The \textsc{Professor} framework~\cite{Buckley:2009bj} introduced surrogate-based parameter optimization, enabling global fits of \pyt parameters to large measurement sets. Machine-learning-assisted approaches such as \textsc{Apprentice}~\cite{Krishnamoorthy:2021nwv} extend this strategy using more flexible surrogate models.

Within this validation and calibration landscape, \pyt commonly serves as a baseline generator in comparative studies. Its parameterized structure supports systematic uncertainty exploration, and its sustained maintenance enables consistent comparisons across experimental eras. Tunes and generator versions are archived and tied to specific analyses, contributing to reproducibility and continuity in phenomenological and experimental work.

\subsubsection{Machine-learning applications}
\label{sec:ml-applications}
A rapidly growing body of work uses \pyt-generated samples as training data for machine-learning models, or wraps around \pyt~to generate such samples on demand. This includes classification studies, generative models that emulate parton showers and hadronization, and inference frameworks. In many such workflows, \pyt acts as the underlying physics simulator from which the statistical structure of the data is learned. Examples of such work includes: \textsc{Homer} \cite{Bierlich:2024xzg}, \textsc{Junipr} \cite{Andreassen:2019txo}, \textsc{OmniLearn} \cite{Mikuni:2025tar}, \textsc{OmniFold} \cite{Andreassen:2019cjw}, \textsc{MadMiner} \cite{Brehmer:2018hga}, \textsc{ParticleNet} \cite{Qu:2019gqs}, \textsc{CNN} \cite{Macaluso:2018tck}, \textsc{ToPoDNN} \cite{Pearkes:2017hku}, \textsc{EnergyFlow} \cite{Komiske:2018cqr} and numerous others. A more detailed discussion is given in Sec.~\ref{sec:ml-hybrid}.

\subsection{Standardization and interface development}
\label{sec:standards}

As described above, \pyt is used in a very large and diverse software ecosystem. 
Such broad integration is only possible because common technical standards have been established and widely adopted within the field. These standards define how events are represented, how particle identities are encoded, how generator-level information is exchanged between codes, and how systematic variations are communicated. 

\pyt has both benefited from and contributed to the development of such standards. In this section we summarize its role in interface standardization within high-energy physics, as well as recent efforts to formalize and structure user contributions through the emerging \texttt{pythia-contrib} framework.

\subsubsection{Development of standards}

The interoperability of event generators, analysis tools, and detector simulation frameworks relies on a shared set of conventions and interfaces. Over the decades, several such standards have emerged, often through close collaboration between generator authors and experimental software groups.

A foundational example is the adoption of the PDG particle numbering scheme \cite{ParticleDataGroup:2024cfk} (chap. 45), which provides a universal identification code for particles across generators and experiments. Closely related is the evolution of event-record formats, from the early \textsc{HEPEVT} \cite{Altarelli:1989hx} standard used at LEP to more modern C++-based representations such as \textsc{HepMC} \cite{Dobbs:2001ck} and \textsc{HepMC3} \cite{Buckley:2019xhk}. These event-record standards allow generator-level information to be passed reliably to detector simulation and analysis frameworks.

For the exchange of hard-scattering information between matrix-element generators and parton-shower programs, the Les Houches Accord (LHA) \cite{Boos:2001cv} and the Les Houches Event File (LHEF) \cite{Alwall:2006yp} formats have played a central role. These standards enable external matrix-element codes to interface consistently with \pyt and other shower generators. Similarly, the SUSY Les Houches Accord (SLHA) \cite{Skands:2003cj} provides a standardized format for the exchange of model parameters in beyond-the-Standard-Model scenarios. These standards are not limited by \pyt~by any means, and are continuously being developed by the community as such. As an example, one could mention recent technical improvements enabling reading and writing LHEF files in HDF5 format \cite{Hoche:2019flt,Bothmann:2023ozs}.

More recently, agreements on the structure and interpretation of event weights \cite{Bothmann:2022pwf} -- including systematic variations for scales, parton distribution functions, and other parameters -- have become increasingly important for large-scale experimental analyses. The implementation of consistent weight-handling mechanisms within \pyt is aligned with these community-wide conventions.

\pyt has not only implemented these standards, but has often been involved in their practical refinement through close interaction with experiments and other generator authors. The existence of shared conventions ensures that external tools can interoperate reliably, that results can be compared across different generators, and that experimental measurements can be reproduced over long time scales.

Standardization also plays a central role in tuning. Global tuning efforts, such as the Monash tune \cite{Skands:2014pea} and experiment-specific tuning campaigns within the LHC experimental collaborations, rely on agreed definitions of observables, parameter blocks, and weight interpretations. Since \pyt implements a complete physics model rather than an isolated component, tuning is not merely a choice of numerical parameters but constitutes a shared calibration layer used across experiments and phenomenological studies. The Monash tune mentioned above has, to date, been cited more than 1500 times. The existence of stable interfaces and documented parameter definitions is therefore essential for preserving the integrity of tuned configurations over time.

Taken together, these standards extend beyond \pyt itself. They enable a diverse ecosystem of external codes to interoperate reliably, and thereby allow results to be compared, validated, and reused across experiments, phenomenology, and adjacent communities.

\subsubsection{User contributions}
\label{sec:pythia-contrib}

The open-source nature of \pyt has historically enabled user contributions at multiple levels. In earlier stages of development, such contributions were often incorporated directly into the main code base, following informal coordination with the authors. Over time, as the code and user community have grown, the need for clearer structure and governance of external contributions has become more pronounced.

To address this, we are developing a more formalized extension framework, \texttt{pythia-contrib}. The goal of this initiative is to provide a standardized and version-controlled environment in which users can develop and distribute physics modules, interfaces, and experimental features that extend \pyt without modifying its core. This structure aims to balance stability of the main release with flexibility for rapid development.

A structured contribution layer offers several advantages. It enables experimental groups and phenomenologists to prototype new models or interfaces in a controlled setting, and it provides clearer attribution and maintenance responsibility for contributed modules. Importantly, it also aligns with the broader ecosystem perspective described in Sec.~\ref{sec:ecosystem}, where modularity and well-defined interfaces are prerequisites for sustainable long-term integration.

In this way, \texttt{pythia-contrib} represents a continuation of the collaborative model that has characterized the development of \pyt from its inception, but adapted to the scale and complexity of the modern high-energy physics software landscape.





\section{Future}\label{sec:future}

The analysis presented in Secs.~\ref{sec:today} and \ref{sec:ecosystem}
demonstrates that \pyt\ has evolved from a modelling tool into a
widely embedded research infrastructure. It supports a large and
diverse user base, operates as a backend component in numerous
simulation frameworks, participates in shared standards, and serves
as a calibration baseline across experimental and phenomenological
studies. This scale of integration implies not only scientific
influence, but institutional responsibility.

The central challenge for the coming decades is therefore not merely
the addition of new physics features, but the sustained development
of a stable, interoperable, and adaptable software facility. As
experimental programmes expand in scale and duration, and as
computational methods evolve, the requirements placed on simulation
infrastructure grow correspondingly. In this section we discuss how
questions of sustainability, experimental evolution, physics
development, emerging computational methods, and hardware trends
intersect in shaping the future of the \pyt facility.

\subsection{Building a sustainable software infrastructure}\label{sec:sustainability}

The sustainability of research software has increasingly become a topic of attention within science policy and infrastructure discussions. In this context, sustainability refers not to environmental impact, but to the long-term availability, maintainability, and institutional anchoring of scientific code and the communities that develop it. International initiatives -- including activities under the High Energy Physics Software Foundation and the EVERSE network \cite{hsf,everse}, as well as recent expert meetings convened by the Organisation for Economic Co-operation and Development (OECD) on access to and sustainability of research software \cite{oecd1,oecd2} -- have highlighted the growing recognition that research software constitutes a critical component of modern scientific infrastructure.

Event generators present particular challenges within this broader landscape. As mentioned in Fig.~\ref{fig:three-prong} and the surrounding discussion, a program such as \pyt~occupies an interdisciplinary space between formal theoretical developments and phenomenological modelling, high-performance software engineering, and tight integration and interaction with experimental collaborations and integration in downstream computational frameworks. Its development requires sustained expertise across these domains, and the accumulation of both physics insight and implementation knowledge over long time scales. Unlike large experimental collaborations with formal governance structures and institutionalised funding streams, generator projects typically operate with comparatively small core teams (see Fig.~\ref{fig:authors}) and distributed contributors, as discussed in Sec.~\ref{sec:operations-model}. This structure enables flexibility and rapid innovation, but also places significant responsibility on a limited number of developers and highlights the importance of continuity in personnel and institutional support.

Sustainability also concerns long-term availability and interpretability. \pyt\ maintains public archives of historical (since 1986) and current releases through its website and version-controlled repositories. Preserving access to earlier versions is essential, as analyses performed within large experimental collaborations may need to be revisited years after publication, often in the context of new theoretical developments or updated statistical interpretations.

Beyond code availability, long-term benchmarking plays an important role. The parallel initiative \textsc{MCplots} \cite{Karneyeu:2013aha} provides continuous comparisons of current and historical generator predictions against experimental data using the \textsc{Rivet} framework. Through distributed volunteer computing resources such as LHC@home \cite{Korneeva:2024oho}, these comparisons are maintained across generator versions and physics tunes. Such efforts contribute to transparency and traceability in the evolution of physics modelling.

Sustainability, however, encompasses more than code maintenance and availability. It includes the cultivation of viable career paths for researchers whose work lies at the boundary between theory, software development and experimental particle physics. Such roles do not always align neatly with traditional disciplinary classifications, yet they are essential for maintaining reliable and interoperable simulation infrastructure.

These considerations become even more pressing in the age of data-intensive science and artificial intelligence. As simulation tools are increasingly embedded in large-scale statistical workflows and ML-based pipelines, expectations regarding reproducibility, interface stability, and transparent configuration management continue to rise. Sustainable research software must therefore be robust not only in its physics content, but also in its governance, documentation, and adaptability to evolving computational paradigms.

In the following subsections, we outline how these sustainability considerations intersect with concrete developments: the demands of upcoming experimental facilities, ongoing improvements in physics modelling, the integration of machine-learning methods, and the implications of emerging hardware architectures.

\subsection{New Experiments}
\label{sec:new-experiments}

The next major operational milestone in collider physics is the High-Luminosity upgrade of the Large Hadron Collider (HL-LHC). The substantial increase in luminosity will be accompanied by a corresponding expansion in simulated event samples required for detector optimisation, background modelling, and precision analyses. This large-scale generation effort is already underway. Approximately half of the computing resources are devoted to detector simulation and event reconstruction, while event generation itself is expected to account for about 20\% of the total computing budget.

A substantial part of this effort concerns the modelling of SM backgrounds to rare or BSM signatures. These simulations often rely on complex multi-leg matrix elements provided by specialised tools such as \textsc{MadGraph}, with \pyt\ supplying parton showering, hadronisation, and underlying-event modelling. The scale of HL-LHC production therefore reinforces the role of \pyt\ as a backend component within a broader ecosystem of interoperable tools. The cumulative computational investment associated with \pyt\ is estimated to be in the millions of CPU-years \cite{ATLAS:2020pnm}, underlining both the scale of deployment and the importance of long-term reliability.

Beyond the LHC programme, the Electron–Ion Collider (EIC) at Brookhaven National Laboratory is currently the only other approved new collider facility. Operating at lower centre-of-mass energies but with unprecedented control over beam polarisation and nuclear structure, the EIC will open new precision studies of the partonic structure of nucleons and nuclei. For \pyt, this programme motivates extensions in the heavy-ion and lepton–hadron sectors, including the consistent treatment of polarised beams and spin correlations. Work in this direction not only supports the EIC physics case, but also enhances the generator's applicability to related environments, such as hadronic cascades in high-energy neutrino experiments (\eg~IceCube and KM3NET).

The ongoing update of the European Strategy for Particle Physics \cite{esppu} is expected to shape the next flagship facility at CERN, with proposals such as the Future Circular Collider (FCC) under active consideration. Such programmes involve multi-decade timelines, with an initial $e^+e^-$ phase dedicated to precision Higgs and electroweak measurements, followed by a later hadron-collider stage \cite{FCC}. The extended time horizons of these projects highlight the importance of durable and adaptable simulation infrastructure. \pyt\ is already extensively used in feasibility studies \cite{FCC}, reflecting not only its modelling capabilities, but also the value of continuity and compatibility across experimental generations.

The modelling of cosmic ray hadronic cascades in the atmosphere is a 
recent addition to \pyt, still in a validation stage \cite{Lonnblad:2025ivw}. 
It requires an administrative framework where it is 
possible to switch between different hadron beams at different 
energies. This is solved by pretabulation of multiparton interaction
and nuclear geometry parameters for about 20 different particle types 
at a grid of energies, with interpolation for the current energy.  
Total, elastic and diffractive cross sections are modelled using 
pomeron plus reggeon ans\"atze at high energy, and attaching to the 
hadronic rescattering expressions at low  energy. Thus it is possible
to trace cascades from the highest observed cosmic-ray energies down to
a few hundred MeV. Parton distribution functions also need to be 
modelled. Two partly separate codes have been developed, which has 
offered useful cross-checks. The \textsc{Angantyr}-based option is 
the more sophisticated one, and the only one that can handle not only 
arbitrary hadron--nucleus collisions, but also nucleus-nucleus ones. 
The \textsc{PythiaCascade} alternative uses a simplified handling of 
nuclear geometry and subsequent collisions in a nucleus, making it 
more transparent and robust, but giving comparable results.
   
This modelling is now being integrated into the \textsc{Corsika~8}
cosmic-ray-tracking framework, which also handles other aspects like 
electromagnetic cascades. Another ongoing integration is with the
detector-simulation program \textsc{Geant~4}, where it would replace
the old Fortran \pyt code currently in use. Future extensions are 
under study, \eg for neutrino or photon beams. 

\subsection{New Physics}
\label{sec:new-physics}

Improvements in physics modelling within \pyt\ are closely tied to developments in experimental precision and to the evolving structure of the generator ecosystem. As measurements become increasingly accurate and statistically powerful, simulation tools must deliver corresponding theoretical reliability while remaining interoperable with specialised external components.

Higher formal precision in hard-scattering processes typically requires next-to-leading order (NLO) or higher matrix elements. In modern workflows, such calculations are often provided by dedicated matrix-element generators such as \textsc{MadGraph}, with \pyt\ supplying parton showering, hadronisation, and underlying-event modelling. This division of labour is not a limitation, but a structural feature of the ecosystem: it allows specialised tools to focus on their respective domains while relying on stable and well-defined interfaces. Maintaining these interfaces, matching procedures, and validation strategies constitutes a significant part of the infrastructure burden, independent of the underlying physics improvements.

For several leading-order processes implemented directly in \pyt, NLO accuracy can alternatively be achieved through internal matching procedures. In these cases, the first emission in the parton shower is constructed to reproduce the real-emission component of the NLO matrix element, while the virtual corrections can be incorporated through multiplicative reweighting factors, in a manner similar to approaches used in \textsc{Powheg}. Such techniques have been implemented for deeply inelastic $ep$ scattering and may be extended to additional processes. Their development illustrates how physics precision and software architecture are intertwined: higher-order accuracy must be achieved without compromising modularity, reproducibility, or compatibility with external tools.

The pursuit of higher perturbative accuracy also places demands on the parton shower itself. Parton showers reorganise the perturbative expansion by resumming logarithmically enhanced terms to all orders, with emissions ordered in an evolution variable that regulates the divergences of the fixed-order expansion. The default shower in \pyt\ is designed for broad applicability and robustness, and formally resums leading logarithmic (LL) contributions. The alternative \textsc{Vincia} shower \cite{Brooks:2020upa} improves the treatment of QCD coherence effects, but remains LL accurate. The ongoing development of the \textsc{Apollo} shower \cite{Preuss:2024vyu}, which systematically incorporates next-to-leading logarithmic (NLL) effects, reflects the increasing demand for formal precision.

The coexistence of multiple shower formulations is scientifically valuable, enabling systematic comparisons and uncertainty estimates. At the same time, it increases the complexity faced by users. Clear documentation, training material, and consistent configuration handling therefore become increasingly essential components of the infrastructure as the amount of coexisting alternatives grows. The availability of multiple precision levels and modelling strategies strengthens the scientific programme, but also reinforces the need for structured governance and user support.

For heavy-ion and high-multiplicity proton collisions, the primary challenges are different. Here the emphasis shifts from formal perturbative precision to the modelling of collective phenomena. In central nucleus–nucleus collisions at the LHC, the formation of a Quark–Gluon Plasma (QGP) is commonly described using hydrodynamic approaches that predict collective flow, strangeness enhancement, and jet quenching (see \eg~ref.~\cite{Adolfsson:2020dhm}). Dedicated QGP event generators implement these mechanisms explicitly and constitute a distinct modelling paradigm.

Within \pyt, the focus has been complementary. A series of models
based on interactions between colour strings and between hadrons
emerging from string fragmentation
\cite{Bierlich:2014xba,Bierlich:2020naj,Sjostrand:2020gyg,Bierlich:2024odg,
  Altmann:2024odn,Altmann:2025afh} have been developed to provide a
microscopic description of collective-like effects without assuming
the formation of a thermalised plasma. These models can be viewed as
providing a realistic null hypothesis in environments where collective
signatures are observed, particularly in small systems where
hydrodynamic interpretations are less straightforward. Their purpose
is not to replace dedicated QGP frameworks, but to offer an
alternative baseline within a unified hadronic modelling framework.

Interoperability plays a central role in this domain. \pyt\ already serves as a backbone for several heavy-ion and QGP-oriented codes (see Table~\ref{table:runtime-integration}), providing \eg hard processes or hadronisation. Studies combining Colour Glass Condensate initial states with string-based hadronisation \cite{Schenke:2016lrs} illustrate how \pyt\ components can be embedded within broader theoretical frameworks. In addition, ongoing developments explore hybrid ``core–corona'' approaches in which \textsc{Angantyr}-based initial states are coupled to external medium descriptions \cite{Kanakubo:2021qcw}. While not yet available as public releases, such developments highlight the technical flexibility of the framework and the importance of maintaining well-defined extension mechanisms, such as those provided by the \texttt{pythia-contrib} structure discussed in Sec.~\ref{sec:pythia-contrib}.

Finally it should be acknowledged that \textit{New Physics} in the
collider community often is assumed to be synonymous to \textit{Beyond
  the Standard Model}, while we here mainly discuss \emph{new physics
  modelling} of the Standard Model. The reason for this is that, from
an event generator perspective, the inclusion of BSM physics signals
normally does not pose very large challenges. Indeed, \pyt does not
include very much BSM physics, and instead relies on external
programs, such as \textsc{MadGraph}, to provide the relevant
production mechanisms. So far, there have not been many hints of BSM
physics at the LHC. Nevertheless, new physics has been discovered,
often resulting in new models being implemented in \pyt. One example
is the string interactions models mentioned above, that were sparked
by the discovery of collective effects in high multiplicity proton
collisions. Another example is the revival of an old framework for above- and
below-threshold production of top pairs \cite{Sjostrand:2025qez},
after recent experimental hints of toponium production.

\subsection{New methods}
\label{sec:ml-hybrid}

Modern collider analyses rely on large simulated event samples not only for central predictions, but also for systematic uncertainty estimates. Traditionally, variations of renormalization and factorization scales, parton distribution functions, multiparton interaction parameters, or hadronization settings required independent event samples for each parameter choice. For large experimental workflows, this quickly becomes computationally prohibitive.

To our knowledge, \pyt was the first general-purpose event generator to implement event-by-event weight variations for parton showers \cite{Mrenna:2016sih}, \ie weights expressing the probability of an already generated shower history under alternative assumptions. This approach has since been generalized to other parts of the event generation procedure, most notably to hadronization models \cite{Bierlich:2023fmh,Assi:2025gog}, providing a unified reweighting framework across perturbative and selected non-perturbative components.

From an infrastructure perspective, this development is central. First, it reduces the computational cost of systematic studies by allowing multiple parameter variations to be evaluated from a single generated sample. Second, it improves reproducibility: alternative predictions are derived from identical underlying event kinematics, eliminating statistical fluctuations between independently generated samples. Third, reweighting can be applied after detector simulation -- typically the most resource-intensive stage -- making it directly compatible with large-scale experimental production chains.

The need for such functionality is reflected in the user-base analysis presented in Sec.~\ref{sec:user-base}. Large LHC collaborations, which dominate in terms of unique users, require efficient and stable uncertainty evaluation across broad physics programs, while phenomenological studies benefit from rapid parameter scans and tuning workflows. The weight-variation framework supports both communities and strengthens \pyts role as a scalable, uncertainty-aware simulation infrastructure. It also opens new opportunities for hybrid and ML-assisted workflows, as mentioned in Sec.~\ref{sec:ml-applications}.

A minor but growing fraction of the \pyt user base, visible in the analysis in Sec.~\ref{sec:user-base}, consists of works at the interface of particle physics and machine learning. These can be roughly divided into two sub-categories: (i) studies where \pyt-generated events are used ``as-is'' as training data for classifiers, and (ii) studies where generative models aim to partially or fully replace or complement \pyt in a given simulation chain as model surrogates.

This shift in usage places new demands on the generator, subtly different for the two cases. 

In the first case, classifier workflows typically require the generation of very large and statistically consistent datasets, clear control over systematic variations
and well-defined interfaces to downstream tools, as outlined in Sec.~\ref{sec:ecosystem}. Maintainence of standardized output formats, weight variations, and robust configuration handling therefore becomes central not only for traditional analyses, but also for ML-based pipelines.

In the second case, generative or surrogate models aim to replace selected components of the simulation chain. Examples include neural-network emulators of parton showers or hadronization, or hybrid workflows where a learned model complements \pyt in specific regions of phase space, as in recent developments such as \textsc{Homer}. In such settings, \pyt serves both as a training target and as a reference against which the surrogate model is validated.

The use of surrogates in generator workflows is not new. Long before the recent surge in machine learning, tools such as \textsc{Professor} implemented fast polynomial-based emulators to accelerate tuning studies. What is new is the scope: modern ML approaches attempt to approximate increasingly large parts of the event-generation chain, sometimes with the ambition of replacing computationally expensive components.

From an infrastructure perspective, this development reinforces rather than diminishes the role of \pyt. Surrogate models require a stable and well-defined reference implementation for training, validation, and uncertainty quantification. Clear configuration handling, reproducible event records, consistent definitions of physics objects, and an open and modular structure (as described in Sec.~\ref{sec:pythia-contrib}) become even more important as approximate models are benchmarked against a trusted baseline.

At the same time, the growing number of surrogate physics modules introduces new technical challenges. Such modules are often formulated in latent spaces and trained against high-level observables produced by \pyt, rather than directly against experimental data. Modifying or replacing one component of the event-generation chain may therefore require correlated adjustments of downstream components, including retuning of non-perturbative parameters. Current tuning workflows are not designed for such tightly coupled, end-to-end optimization. One of the long-term motivations for reweightable simulations is precisely to enable more continuous and eventually differentiable pipelines, opening the possibility of consistent ML–MC hybrid approaches.

Taken together, these two modes of ML usage illustrate how \pyt functions both as a large-scale data generator and as a reference standard. We view these developments as a natural and welcome extension of the generator ecosystem, providing valuable feedback on both performance and physics modelling. Maintaining this dual role is essential if \pyt is to continue serving as a common foundation across experimental analyses, phenomenology, and emerging computational paradigms.

\subsection{New Hardware}
\label{sec:new-hardware}

As Moores Law nowadays seems to be broken, and the processors no
longer double in speed every other year, more focus is necessarily put
on increasing efficiency of computer programs restructuring the code
to make use of parallelisation. For event generation, this is not
trivially achieved. Most of the generation of events in \pyt is
inherently sequential, where each step depends on everything that has
happened before. Currently the only way of parallelising \pyt is
therefore to use multi-processor CPUs and simply run several \pyt
instances in parallel threads.

There may be some speed to be gained with modern SIMD (Single
Instruction, Multiple Data) CPUs, and this will be explored in the
future. Also the ML techniques described in the previous subsection
can obviously benefit from GPU processors, but for the normal event
generation in \pyt, the benefit is more uncertain. Preliminary studies
have, however, found that for the simulation of collective effects (see
section \ref{sec:new-physics}), it may be a case for using the massive
parallel powers of GPU processors. The issue here is that the
complexity of e.g.\ the hadronic rescattering modelling grows as the
square of the number hadrons produced, and in lead collisions at the
LHC, these count in tens of thousands. Even if the checking of whether
a given pair of hadrons may interact with each other is very quick, if
you need to do that a hundred million times for each event, it soon
gets forbiddingly slow. Due to the rather slow communication between
the CPU and the GPU, preliminary test only give a modest overall
speed-up of a factor 3. But this is certainly something that will be
pursued further, also for the other models for collective effects.

Further into the future one may expect that quantum computers will
become commonplace. There have been a couple of suggestions for how to
use quantum computing for event generators, e.g.\ for parton showers
\cite{Gustafson:2022dsq} and for string fragmentation \cite{Yannick},
but there are currently no active plans to pursue this for \pyt.

\section{Conclusion}
\label{sec:conclusions}

This contribution has argued that \pyt\ is more than a Monte Carlo event
generator: it is a widely embedded research infrastructure. Its
integration into experimental production chains, its role in
standardisation and validation frameworks, and the scale of its user
community demonstrate that it functions as a shared virtual-reality facility for
high-energy physics and beyond. Its value lies not only in the physics
models it implements, but in the stability and interoperability it
provides.

Behind the figures of ``tens of thousands of users'' are individual
researchers whose daily work depends on reliable and transparent
simulation tools: PhD students completing analyses under tight
deadlines, postdoctoral researchers testing new ideas, and principal
investigators relying on open standards for communication within large
collaborations. For these users, \pyt\ is an enabler -- a framework
through which hypotheses are confronted with data, uncertainties are
quantified, and results become reproducible.

This role is visible not only in citation metrics, but in continuous
interaction with the community through issue trackers, helpdesks,
summer schools, and conferences. In these exchanges, infrastructure
becomes tangible: the code provides functionality, and the community
gives it purpose.

Infrastructure, however, does not sustain itself. A comparatively small
group of developers maintains and extends the framework while ensuring
stability and backward compatibility. As illustrated in
Fig.~\ref{fig:three-prong}, the development of \pyt\ requires expertise
spanning theory, phenomenology, software architecture, and experimental
practice accumulated over long time scales and not easily
replaced.

Such work does not fit neatly within traditional academic categories.
In funding environments that emphasise short-term projects and
discipline-specific evaluation metrics, sustaining long-lived,
interdisciplinary software efforts presents structural challenges.
Supporting viable career paths in this space is therefore essential for
scientific continuity.

The developers of \pyt\ act as stewards of a shared scientific
resource. Their task is not only to implement new physics ideas, but to
preserve accumulated knowledge, maintain interoperability, document
standards, and train new users -- work that is most effective when it
remains closely coupled to the community it serves.

The scientific landscape continues to evolve. New experimental
facilities increase demands on precision and scale, while advances in
statistical inference and machine learning reshape analysis workflows.
These developments heighten the need for modularity, transparent
interfaces, and stable reference implementations. By remaining
interoperable and supporting inference-oriented approaches, \pyt\
adapts to emerging paradigms while preserving the continuity required
by large collaborations.

Ultimately, \pyt\ exists to \textit{enable people to do science}. Its future
depends as much on the strength of its community and the support of its
stewards as on advances in physics or computing. Sustaining that balance
keeps the triangle of theory, software, and experiment intact, ensuring
that this shared infrastructure continues to serve future generations of
scientists.

\backmatter







\bibliography{Pythia}

\end{document}